\documentclass[12pt]{article}
\usepackage{graphics}
\usepackage{epsfig}
\usepackage{amsmath}

\begin{document}

\setlength{\textheight}{240mm}
\voffset=-15mm
\baselineskip=20pt plus 2pt
\renewcommand{\arraystretch}{1.6}

\begin{center}

{\large \bf The Relation Between the Quasi-localized Energy Complexes
and the Thermodynamic Potential  for the  Schwarzschild-de Sitter Black Hole}\\
\vspace{5mm}
\vspace{5mm}
I-Ching Yang  \footnote{E-mail:icyang@nttu.edu.tw}

Department of Applied Science, National Taitung University, \\
Taitung 95002, Taiwan (R.O.C.)\\

\end{center}
\vspace{5mm}

\begin{center}
{\bf ABSTRACT}
\end{center}
The Schwarzschild-de Sitter (SdS) black hole solution, which has two event horizons, is considered 
to examine the relation between the quasi-localized energy complexes on ${\cal M}$ and the heat 
flows passing through its boundary $\partial {\cal M}$.  Here ${\cal M}$ is the patch between 
cosmological event horizon and black hole event horizon of the SdS black hole solution.  Conclusively, 
the relation, like the Legendre transformation, between the quasi-localized Einstein and M{\o}ller 
energy complex and the heat flows passing through the boundary is obeyed, and these two
quasi-localized energy complexes could be corresponding to thermodynamic potentials.

\vspace{2cm}
\noindent
{PACS No.:04.60.Cf, 04.70.Dy.  \\}

\vspace{5mm}
\noindent

\newpage

Recently, Yang et al.~\cite{Y12, YCT12,YH14} have inferred that the formula about the quasi-localized Einstein and 
M{\o}ller energy complexes on ${\cal M}^*$ and the heat flows passing through the boundary of ${\cal M}^*$, like as
\begin{equation} 
\left. E_{\rm E} \right|_{{\cal M}^*} = \left. E_{\rm M} \right|_{{\cal M}^*} - \sum_{\partial {\cal M}^*} {\bf TS}  .
\end{equation}
Here ${\cal M}^*$ is the patch between event horizon ${\cal H_+}$ located at $ r= r_+$ and  inner Cauchy horizon 
${\cal H_-}$ located at $ r= r_-$ for  the spherically symmetric black hole with two separate horizons.  Eq. (1) is similar
to the Legendre transformation between Helmholtz free energy $F$ and internal energy $U$ or between Gibbs free 
energy $G$ and enthalpy $H$, and therefore these quasi-localized pseudotensor energy complexes 
$\left. E_{\rm E} \right|_{{\cal M}^*}$ and $\left. E_{\rm M} \right|_{{\cal M}^*}$ could correspond to 
thermodynamic potentials.  But, in previous studies, $r_-$ is not the event horizon, in which has an acceptable definition 
for the temperature and entropy of black hole.  In this article, I would like to consider the Schwarzschild-de Sitter (SdS) 
black hole solution, which has two separate event horizons, and to review the relation  between  the quasi-localized 
pseudotensor energy complexes on a patch between two event horizons and the heat flows passing through its boundary.

The SdS black hole solution~\cite{K18}, describing a spherically symmetric solution of the vaccum Einstein field 
equations in the presence of a positive cosmological constant $\Lambda$
\begin{equation}
R_{\mu \nu} -\frac{1}{2} g_{\mu \nu} R +\Lambda g_{\mu \nu} = 0  ,
\end{equation}
is given in the static form
\begin{equation}
ds^2 = f(r) dt^2 -f^{-1} (r) dr^2 -r^2 d\Omega 
\end{equation}
and its metric function was solved as 
\begin{equation}
f(r) = 1- \frac{2M}{r} - \frac{r^2}{b^2}  ,
\end{equation}
where $b^2 = 3 / \Lambda$.
Here, we shall consider the metric function in a factorization form 
\begin{equation}
f(r)= -\frac{1}{r b^2} (r-r_c)(r-r_b)(r-r_0)  .
\end{equation}
While $ 3\sqrt{3} M/b < 1$ , the metric function would has two distinct positive real roots $r_c$ and $r_b$, and the
smaller one $r_b$ and the larger one $r_c$ would be regarded as the position of the black hole event horizon and 
the cosmological event horizon for observers moving on the world lines of constant $r$ between $r_b$ and $r_c$.
Compared with Eq. (4), the relations of these three roots are given by  
\begin{subequations}
\begin{align}
& r_c + r_b + r_0 = 0  , \\
& r_c r_b + r_c r_0 + r_b r_0 = - b^2  , \\
& r_c r_b r_0 = - 2M b^2  .
\end{align}
\end{subequations}
Because of Eq. (6a), $r_0 = -(r_c +r_b )$, the metric function is taken to be
\begin{equation}
f(r) = -\frac{1}{r b^2} (r-r_c)(r-r_b)(r+r_c +r_b)  ,
\end{equation}
and Eq. (6b) and Eq. (6c) are also reorganized as 
\begin{subequations}
\begin{align}
& r_c^2 +r_c r_b + r_b^2 = b^2 ; \\
& r_c r_b (r_c + r_b ) = 2M b^2  .
\end{align}
\end{subequations}
Let ${\cal S}^2 (r)$ be a 2-sphere of radius $r$.  Thus we shall suggest that 
${\cal M} = \left\{ {\cal S}^2 (r) | r_c > r > r_b \right\} $ is the patch between cosmological event 
horizon ${\cal H}_C ={\cal S}^2 (r_c)$ and black hole event horizon ${\cal H}_B = {\cal S}^2 (r_b)$,
and the boundary of ${\cal M}$ is $\partial {\cal M} = {\cal H}_C \cup {\cal H}_B$.  The patch 
${\cal M}$ is the region I of Penrose diagram for SdS black hole solution (shown as Fig. 1).
\begin{figure}[h]
  \centering
  \includegraphics[width=0.8\textwidth]{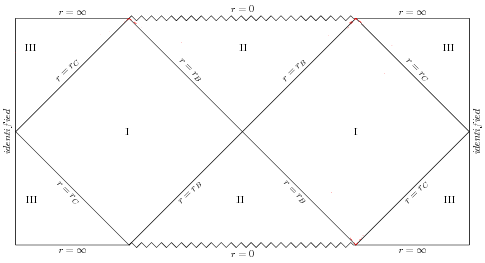} 
  \caption{Penrose diagram for SdS black hole solution.}
\end{figure}

In the study of the thermodynamics of SdS black hole by Gibbons and Hawking~\cite{GH77}, the 
Hawking temperature ${\bf T}$~\cite{V} of ${\cal H}_C$ and ${\cal H}_B$ are 
\begin{subequations}
\begin{align}
\left. {\bf T} \right|_{\cal H_C} & = \frac{1}{4\pi r_c b^2} (r_c - r_b)(2 r_c + r_b) , \\
\left. {\bf T} \right|_{\cal H_B} & = \frac{1}{4\pi r_b b^2} (r_c - r_b)(r_c + 2 r_b) ,
\end{align}
\end{subequations}
and the Bekenstein-Hawking entropy ${\bf S}$~\cite{H74,H75}  of ${\cal H}_C$ and ${\cal H}_B$ are
\begin{subequations}
\begin{align}
\left. {\bf S} \right|_{\cal H_C} & = \pi r_c^2 , \\
\left. {\bf S} \right|_{\cal H_B} & = \pi r_b^2  .
\end{align}
\end{subequations}
For those two event horizons, the heat flowa are evaluated as  
\begin{subequations}
\begin{align}
\left. {\bf TS} \right|_{\cal H_C} & = \frac{1}{4 b^2} (r_c - r_b) (2r_c^2 + r_c r_b) , \\
\left. {\bf TS} \right|_{\cal H_B} & = \frac{1}{4 b^2} (r_c - r_b) (r_c r_b + 2r_b^2) .
\end{align}
\end{subequations}
Hence, the heat flow passing through the boundary $\partial {\cal M}$ would be expressed by
\begin{equation}
\left. {\bf TS} \right|_{\partial {\cal M}} =\frac{1}{2 b^2} (r_c^3 -r_b^3) .
\end{equation}

Subsequently, on ${\cal M}$, the quasi-localized energy complexes  in the Einstein~\cite{T,E} ,
and M{\o}ller~\cite{M58,M61} prescription should be considered.  The energy component of 
the Einstein energy-momentum complex~\cite{T,E} is given by
\begin{equation}
E_{\rm E} = \frac{1}{16\pi} \oint H_0^{\;\;0i} \hat{n}_i \cdot d\vec{S} ,
\end{equation}
where 
\begin{equation}
H_0^{\;\;0i} = \frac{g_{00}}{\sqrt{-g}} \frac{\partial}{\partial x^m} \left[ (-g) g^{00} g^{im} \right] 
\end{equation}
and $\hat{n}_i$ is the outward unit normal vector over the infinitesimal surface element 
$d\vec{S}$.  The energy component within radius $r$ obtained by the Einstein complex is 
\begin{equation}
E_{\rm E} = \frac{r}{2} (1 - f)  .
\end{equation}
Therefore, the quasi-localizes Einstein energy complex on ${\cal M}$ is
\begin{eqnarray}
\left. E_{\rm E} \right|^{r_c}_{r_b} & = & \frac{1}{2} (r_c - r_b)  \nonumber \\
& = & \frac{1}{2 b^2} (r_c^3 - r_b^3 )
\end{eqnarray}
Moreover, according to the definition of the M{\o}ller energy complex~\cite{M58,M61} and 
Gauss's theorem, the energy component is given as 
\begin{equation}
E_{\rm M} = \frac{1}{8\pi} \oint \chi_0^{\;\;0i} \hat{n}_i \cdot d\vec{S} ,
\end{equation}
where 
\begin{equation}
\chi^{0i}_0 =  \sqrt{-g} \left( - \frac{\partial g_{00}}{\partial x^i} \right) g^{00} g^{ii}  .
\end{equation}
So the energy component with radius $r$ obtained using the M{\o}ller complex is
\begin{eqnarray}
E_{\rm M} & = & \frac{r^2}{2} \frac{df}{dr} \nonumber  \\
& = & -\frac{r}{2} f - \frac{r}{2 b^2} \left[ (r - r_c )(r + r_b + r_c )  \right. \nonumber \\
& & \left. + (r - r_b )(r + r_b +r_c) + (r-r_b )(r-r_c )  \right], 
\end{eqnarray}
and the quasi-localizes M{\o}ller energy complex on ${\cal M}$ is
\begin{eqnarray}
\left. E_{\rm M} \right|^{r_c}_{r_b} & = & - \frac{1}{2 b^2} 
[ r_c (r_c - r_b )(2 r_c + r_b ) - r_b (r_b - r_c )(r_c +2 r_b ) ]  \nonumber \\
& = & - \frac{1}{b^2} ( r_c^3 - r_b^3) 
\end{eqnarray}

Consequently, the difference of energy between the Einstein and M{\o}ller 
prescription~\cite{YR} is defined as 
\begin{equation}
\Delta E = E_{\rm E} - E_{\rm M}  .
\end{equation}
According to Eq.(16) and Eq.(20), the difference of energy in the patch ${\cal M}$ is 
\begin{equation}
\left. \Delta E \right|_{\cal M} = \frac{3}{2 b^2} ( r_c^3 - r_b^3)  ,
\end{equation}
and its value is triple of the heat flow passing through the boundary $\partial {\cal M}$ 
\begin{equation}
\left. \Delta E \right|_{\cal M} =  3  {\bf TS} |_{\partial {\cal M}}   .
\end{equation}
In this way, the quasi-localized Einstein energy complex 
$\left. E_{\rm E} \right|_{\cal M}$ and M{\o}ller energy complex
$\left. E_{\rm M} \right|_{\cal M}$ will combine with the heat flow passing 
through the boundary $\partial {\cal M}$ 
\begin{equation}
\left. E_{\rm E} \right|_{\cal M} = \left. E _{\rm M} \right|_{\cal M}
+ 3 {\bf TS} |_{\partial {\cal M}}  ,
\end{equation}
although the factor $`` +3 "$ of the heat flow passing through $\partial {\cal M}$ 
in Eq.(24) is different from the factor  $`` -1 "$ in Eq.(1).  As a matter of fact,
$\left. E _{\rm M} \right|_{\cal M} $ must be positive if ${\cal M}$ is dominated by 
attactive gravitation.  For that reason, I prefer that $ \left. E _{\rm M} \right|_{\cal M} $ 
replace by its absolute value $ \left|  \left. E _{\rm M} \right|_{\cal M} \right|$. 
Finally, the difference of energy between the Einstein and M{\o}ller prescription on 
the patch ${\cal M}$ is equal to the heat flow passing by its boundary 
$\partial {\cal M}$, as the formula previously pointed out~\cite{Y12}
\begin{equation}
\left. E_{\rm E} \right|_{\cal M} = \left|  \left. E _{\rm M} \right|_{\cal M} \right| 
- \sum_{\partial {\cal M}} {\bf TS}  .
\end{equation}
Because all boundaries of ${\cal M}$ are event horizons, the summation of the heat flows 
passing through those boundaries $ \sum_{\partial {\cal M}} {\bf TS}$ is will-defined.  In 
conclusion, for the SdS black hole solution, the establishment of Legendre transformation 
in Eq. (25) exhibits that $\left. E_{\rm E} \right|_{\cal M}$  and 
$ \left|  \left. E _{\rm M} \right|_{\cal M} \right| $ would play the role of  
thermodynamic potential.  It conforms with the viewpoint of Nester et al.~\cite{CNC}
and our latest studies~\cite{Y12, YCT12, YH14}. 

\vspace{10mm}
{\bf ACKNOWLEDGMENTS} \\
I would like to thanks Prof. Ching-Tang Tsao for useful suggestions.  This work was 
partially supported by the Ministry of Science and Technology (Taiwan, R.O.C.) under 
Contract No. MOST103-2633-M143-001.

\end{document}